\documentclass[11pt]{article}
\usepackage{a4wide,epsfig,amsmath,amssymb,cite,scalefnt}
\newcommand{\lae}{\lower 2pt \hbox{$\, \buildrel {\scriptstyle <}
\over {\scriptstyle
\sim}\,$}}
\newcommand{\gae}{\lower 2pt \hbox{$\, \buildrel {\scriptstyle >}
\over {\scriptstyle
\sim}\,$}}

\newcommand{\susy}{{\abbrev SUSY}}
\newcommand{\sm}{{\abbrev SM}}
\newcommand{\mssm}{{\abbrev MSSM}}
\newcommand{\amsb}{{\abbrev AM}}
\newcommand{\msugra}{{\abbrev MSUGRA}}
\newcommand{\cmssm}{{\abbrev CMSSM}}

\newcommand{\RG}{{\abbrev RG}}
\newcommand{\abbrev}{\scalefont{.9}}
\def\besub{\begin{subequations}}
\def\eesub{\end{subequations}}
\newcommand{\fig}[1]{Fig.~\ref{#1}}
\newcommand{\reference}[1]{Ref.~\cite{#1}}
\newcommand{\eqn}[1]{Eq.~(\ref{#1})}
\def \in{\leftskip = 40 pt\rightskip = 40pt}

\def \out{\leftskip = 0 pt\rightskip = 0pt}

\def\be{\begin{equation}}
\def\ee{\end{equation}}
\def\bea{\begin{eqnarray}}
\def\eea{\end{eqnarray}}
\def\Tr{{\rm Tr }}
\def\mbar{{\overline{m}}}

\def\frak#1#2{{\textstyle{\frac{#1}{#2}}}}

\def\btil{\tilde b}  
  
\def\dtil{\tilde d}  
\def\etil{\tilde e}   
\def\gtil{\tilde g}

\def\ttil{\tilde t}
\def\util{\tilde u}

\def\tautil{\tilde \tau}

\def\nutil{\tilde \nu}

\def\thbar{\bar\theta}

\def\GeV{\hbox{GeV}}
\def\TeV{\hbox{TeV}}
\def\half{\frac{1}{2}}
\def\lf{16\pi^2}
\def\llf{(16\pi^2)^2}

\def\nn{\nonumber\\}
\def\DRED{\ifmmode{{\rm DRED}} \else{{DRED}} \fi}
\def\DREDD{\ifmmode{{\rm DRED}'} \else{${\rm DRED}'$} \fi}
\def\NSVZ{\ifmmode{{\rm NSVZ}} \else{{NSVZ}} \fi}
\def\Ycal{{\cal Y}}
\def\Gcal{{\cal G}}
  
\def\pa{\partial}

\def\sic{supersymmetric}

\def\Hbar{{\overline{H}}}

\def\fivebar{{\overline{5}}}

\def\bxhat{\hat\beta_{\xi}}

\def\sy{supersymmetry}
\def\sic{supersymmetric}

\def\semi{;\hfil\break}

\begin{document}
\begin{titlepage}
\begin{flushright}
LTH 757\\
\end{flushright}

\vspace*{3mm}

\begin{center}
{\Huge
Anomaly Mediation, Fayet-Iliopoulos\\ 
$D$-terms and the Renormalisation\\
\vskip 0.4cm
Group}\\[12mm]

{\bf R.~Hodgson, I.~Jack, D.R.T.~Jones}\\
%\end{center}   

\vspace{5mm} 
Dept. of Mathematical Sciences,
University of Liverpool, Liverpool L69 3BX, UK\\
\vspace{8mm}

\vspace{3mm}
\end{center}

\begin{abstract}

We address renormalisation group evolution issues that arise  in the
Anomaly Mediated Supersymmetry Breaking scenario when the tachyonic
slepton problem is resolved by Fayet-Iliopoulos term contributions. We
present typical sparticle spectra both  for the original formulation of
this idea and an alternative using  Fayet-Iliopoulos terms for a $U_1$
compatible with a  straightforward GUT embedding.

\end{abstract}

\vfill

\end{titlepage}

%%%%%%%%%%%%%%%%%%%%%%%%%%%%%%%%%%%%%%%%%%%%%
\section{Introduction}

Anomaly mediation (\amsb)~\cite{lrrs}-\cite{amsbtwo} as the 
main source of \sy\ breaking is an attractive idea. In \amsb, 
the soft \sy-breaking
$\phi^*\phi$ masses, $\phi^3$ couplings and gaugino masses 
are all determined by  the appropriate power of the gravitino mass 
multiplied by perturbatively calculable functions 
of the dimensionless couplings of the underlying \sic\ theory. 
Moreover these  functions are \RG{} invariant; that 
is, their renormalisation scale dependence is correctly
given by the renormalisation scale dependence of the dimensionless 
couplings. To put it another way, 
the \amsb{} predictions are UV-insensitive~\cite{Arkani-Hamed:2000xj}.

In recent papers  we have explored a specific version of \amsb, where the
tachyonic  slepton problem characteristic of a minimal implementation of
\amsb{}  is solved by means of an additional $U_1$ gauge symmetry,
$U_1^{\prime}$,   that is broken at high energies. The scale of this
breaking  may be set by a Fayet-Iliopoulos (FI) 
$D$-term~\cite{amsbone}  or
via dimensional transmutation~\cite{amsbtwo}.  In the former  case we 
showed how it is quite natural for the effects of $U_1^{\prime}$ to
decouple at low energies apart from contributions to the scalar masses,
of the form of $U_1^{\prime}$ FI terms, 
which are automatically of the same order as the \amsb{} ones. In the
latter  we argued that it was possible to dispense with an explicit FI
term,  generating the $U_1^{\prime}$ breaking scale via dimensional 
transmutation, exploiting a flat $D$-term direction.  In our explicit
model,  the low energy theory consisted of the usual \mssm{} fields with
 an additional gauge singlet chiral supermultiplet which is  weakly
coupled  to the \mssm{} fields\footnote{The existence of this light 
field is in fact a consequence of general arguments concerning  \amsb{}
decoupling given by Pomarol and Rattazzi~\cite{aprr}.}; the possible cosmological
and phenomenological implications (in particular the possibility  that
its fermionic component might be the LSP) remain to be discussed.  In
this paper we will confine ourselves to the first possibility, where the
 low energy theory simply consists of the \mssm{} fields.  

In both the above scenarios, there is, however, a subtlety  with regard
to the afore-mentioned \RG{} invariance, concerning the 
Fayet-Iliopoulos term associated with the \sm{} (or \mssm) $U_1$, 
$U_1^{\sm}$.  Suppose for simplicity  the $U_1^{\prime}$ breaking scale
coincides with  the gauge unification scale $M_X$, and that the
$U_1^{SM}$ FI term is zero there. It turns out that the presence of the
$U_1^{\prime}$ FI terms   in the effective  field theory means that even
though it is zero at $M_X$,  the $U_1^{SM}$ FI term can become
significant in the evolution to low energies. Thus there will be
contributions of FI form  for both the $U_1^{\prime}$ and the $U_1^{SM}$
to the scalar masses.  Now as emphasised in \reference{mwells}, these
two contributions can be reparametrised into a contribution of  the form
of a single $U_1^{\prime\prime}$ FI-term.  It should now be clear,
however, that the resulting form of this contribution will
be a function of scale since the  size of the $U_1^{SM}$ FI term
generated is a function of scale. 

The upshot is that if we choose a $U_1^{\prime}$ with 
charges for the lepton doublets and singlets chosen so as to solve  the
tachyonic slepton problem, and also zero FI term for  $U_1^{SM}$  at
$M_X$, the resulting spectrum will correspond to a  nonzero FI term for
$U_1^{SM}$ at $M_Z$, {\it or\/}  a zero FI term for $U_1^{SM}$ at $M_Z$
with a different pair of $U_1^{\prime}$ leptonic charges.  

In this paper we shall firstly explain this issue in some detail  and
then repeat some of the precision calculations of Ref.~\cite{amsbone}
but now imposing boundary conditions at $M_X$, and  taking the
opportunity to update input values and correct  some minor bugs in our
previous analysis. 

In the second part of the paper we consider a variation of the same idea
where we augment the theory in a minimal way so as  to render the
$U_1^{\prime}$ charge assignments compatible with a GUT embedding; 
specifically $SU_5$, $SO_{10}$ or $E_6$. Even with the assumption  that
the low energy theory below $M_X$ consists only of the \mssm{} fields,
the resulting allowed region for the
leptonic charges and the sparticle spectrum is quite different from the previous case. 

We also derive some mass sum rules independent of the $U_1^{\prime}$ charges for 
this case, similar to the sum rules given in Refs.~\cite{jja,amsbone}.

\section{The General Case}
First of all, for completeness and to establish notation,  
let us recapitulate some standard results. 
We take an $N=1$ supersymmetric gauge theory with gauge group
$\Pi_{\alpha} G_{\alpha}$
and with superpotential
\be
W(\Phi)=\frak{1}{6}Y^{ijk}\Phi_i\Phi_j\Phi_k+
\frak{1}{2}\mu^{ij}\Phi_i\Phi_j.
\label{eqf}
\ee
We also include the standard soft supersymmetry-breaking terms
\be
L_{\rm SB}=-(m^2)^j_i\phi^{i}\phi_j-
\left(\frak{1}{6}h^{ijk}\phi_i\phi_j\phi_k+\frak{1}{2}b^{ij}\phi_i\phi_j
+ \frak{1}{2}M\lambda\lambda+{\rm h.c.}\right)
\label{Aaf}
\ee
where $\phi^i = (\phi_i)^*$.

For the moment let us assume that the gauge group has one abelian
factor,  which we shall take to be $G_1$. We shall denote  the
hypercharge matrix for $G_1$ by  $\Ycal^i{}_j = \Ycal^j \delta^i{}_j$
and its gauge coupling by $g_1$. 

At one loop we have
\besub
\bea
\label{beone}
\lf\beta_{g_{\alpha}}^{(1)} &=& g_{\alpha}^3Q_{\alpha} =
g_{\alpha}^3\left[T(R_{\alpha})-3C(G_{\alpha})\right], \label{beone:1}\\
\lf\gamma^{(1)i}{}_j &=& P^i{}_j
=\frak{1}{2}Y^{ikl}Y_{jkl}-2\sum_{\alpha}g_{\alpha}^2[C(R_{\alpha})]^i{}_j.
\label{beone:2}
\eea
\eesub
Here $ \beta_{g_{\alpha}}$ are the  gauge $\beta$-functions and $\gamma$ is the  
chiral supermultiplet anomalous dimension, 
$R_{\alpha}$ is the group representation for $G_{\alpha}$ acting on the
chiral fields,  $C(R_{\alpha})$ the corresponding quadratic Casimir and
$T(R_{\alpha}) = (r_{\alpha})^{-1}\Tr [C(R_{\alpha})]$ ,
$r_{\alpha}$ being  the dimension of $G_{\alpha}$. For the adjoint
representation, $C(R_{\alpha})=C(G_{\alpha})I_{\alpha}$, where $I_{\alpha}$ is
the $r_{\alpha}\times r_{\alpha}$ unit matrix.
Obviously 
$T(R_1)=\Tr[\Ycal^2]$, $\left[C(R_1)\right]^i{}_j=(\Ycal^2)^i{}_j$ and 
$C(G_1) = 0$. 
At two loops we have
\bea
\label{betwo}
\llf\beta_{g_{\alpha}}^{(2)}&=&2g_{\alpha}^5C(G_{\alpha})Q_{\alpha}
-2g_{\alpha}^3r_{\alpha}^{-1}\Tr\left[P C(R_{\alpha})\right],
\label{betwo:1}\\
\llf\gamma^{(2)i}{}_j&=& 
2\sum_{\alpha}g_{\alpha}^4C(R_{\alpha})^i{}_jQ_{\alpha}
-\left[Y_{jmn}Y^{mpi}
+2\sum_{\alpha}g_{\alpha}^2
C(R_{\alpha})^p{}_j\delta^i{}_n\right]P^n{}_p.
\label{betwo:2}
\eea

The one-loop $\beta$-functions
for the soft-breaking couplings are given by
\besub
\bea
\lf\beta_h^{(1)ijk} &=&U^{ijk}+U^{kij}+U^{jki},\\
\lf\beta_b^{(1)ij} &=&V^{ij}+V^{ji},\\
\lf[\beta_{m^2}^{(1)}]^i{}_j &=&W^i{}_j,\label{eq:bsoftc}\\
\lf\beta^{(1)}_{M_{\alpha}} &=&2g^2_{\alpha}Q_{\alpha}M_{\alpha},
\eea
\eesub
where
\bea
U^{ijk} &=&h^{ijl}P^k{}_l+Y^{ijl}X^k{}_l,\nn
V^{ij} &=&b^{il}P^j{}_l+\frak{1}{2}Y^{ijl}Y_{lmn}b^{mn}
+\mu^{il}X^j{}_l,\nn
W^j{}_i &=&\frak{1}{2}Y_{ipq}Y^{pqn}(m^2)^j{}_n
+\frak{1}{2}Y^{jpq}Y_{pqn}(m^2)^n{}_i
+2Y_{ipq}Y^{jpr}(m^2)^q{}_r\nn & & \quad
+h_{ipq}h^{jpq}-8\sum_{\alpha} g_{\alpha}^2M_{\alpha}M_{\alpha}^*C(R_{\alpha})^j{}_i,
\eea
with
\be
X^i{}_j=h^{ikl}Y_{jkl}
+4\sum_{\alpha} g_{\alpha}^2M_{\alpha}C(R_{\alpha})^i{}_j.
\ee
We have excluded from \eqn{eq:bsoftc}\ a $D$-tadpole contribution  which
arises if we calculate with the auxiliary field $D$ eliminated.  
If we work in the $D$-eliminated form of the theory then we have instead 
of \eqn{eq:bsoftc}:
\be
\lf[\beta_{m^2}^{(1)}]^i{}_j \to W^i{}_j 
+ 2g^2\Ycal^i{}_j\Tr[\Ycal m^2].
\label{eq:delim}
\ee
This extra 
contribution is only nonvanishing in a theory whose gauge group has an
abelian factor. It can be 
equivalently viewed as a renormalisation of the Fayet-Iliopoulos
parameter,  as we shall now describe. 

In $N=1$ supersymmetric gauge theories whose gauge group has an 
abelian factor,
there exists a possible
invariant that is not otherwise allowed: the
Fayet-Iliopoulos $D$-term,
\be
L =
\xi\int V (x,\theta, \thbar)\,d^4\theta  = \xi D(x).
\label{dta}
\ee

The significance of the $\xi$ term is of course well known. 
The part of the 
scalar potential dependent on the $U_1$ $D$-field is 
\be
V_D  = -\frak{1}{2}D^2 - D\left(\xi + g_1 \phi_i \Ycal^i{}_j \phi^j\right),
\ee
which upon elimination of  the 
auxiliary field $D$ becomes
\be
V_D  = \frak{1}{2}(\xi + g_1 \phi_i \Ycal^i{}_j \phi^j)^2,
\label{eq:dpot}
\ee
so that to obtain a \sic\ ground state we require  at least one field
$\phi^i$ to have a charge  with the opposite sign to $\xi$, and to
develop a vacuum expectation value. Thus  for \sy\ to be unbroken on the
scale set by $\xi$ it is necessarily the case  that the corresponding
$U_1$ is spontaneously broken.  In \reference{amsbone}\ we showed that
in the presence of anomaly mediation soft \sy-breaking terms 
it is quite natural  for the $U_1$ symmetry to be  broken at a
large scale characterised by $\xi$ while all scalars receive, 
from the $U_1$ $D$-term, $(\hbox{mass})^2$ contributions 
characterised  by the gravitino (or anomaly mediation) mass. 

In previous papers \cite{xione}--\cite{xithree} we have discussed the
renormalisation of $\xi$ in the presence of the soft terms.
The result for $\beta_{\xi}$ is as follows:
\be
\beta_{\xi} = \frac{\beta_g}{g}\xi + \bxhat
\label{exacta}
\ee
where $\bxhat$ is determined by $V$-tadpole (or in components
$D$-tadpole) graphs, and is independent of $\xi$.

We found that 
\be
\lf\bxhat^{(1)} = 2g_1\Tr\left[\Ycal m^2\right], \label{eq:bxione}
\ee
\be
\lf\bxhat^{(2)}=-4g_1\Tr\left[\Ycal m^2 \gamma^{(1)}\right].\label{eq:bxitwo}
\ee
The three-loop contribution was computed in \reference{xitwo} for 
an abelian theory and for the \mssm{} in Ref.~\cite{xithree}.

\section{\label{sec:amsbsol}The AM Solution}

Remarkably the following results are \RG{} invariant~\cite{jjb}:
\besub
\bea
M_{\alpha} & = & m_0 \beta_{g_{\alpha}}/g_{{\alpha}},\label{eq:ama}\\
h^{ijk}& = & -m_0\beta^{ijk}_{Y},\\
(m^2)^i{}_j & = & \frac{1}{2}m_0^2\mu\frac{d}{d\mu}\gamma^i{}_j,\label{eq:amc}\\
b^{ij} & = & \kappa m_0 \mu^{ij} - m_0 \beta_{\mu}^{ij}.
\label{eq:amd}
\eea
\eesub
Here $\beta_Y$ is the  Yukawa 
$\beta$-function, given by 
\be
\beta^{ijk}_{Y} = \gamma^i{}_l Y^{ljk} +  \gamma^j{}_l Y^{ilk} +
\gamma^k{}_l Y^{ijl},
\ee
with a similar expression for $\beta_{\mu}^{ij}$. 
It must be emphasised that the \RG{} invariance of 
\eqn{eq:amc}\ holds in the $D$-uneliminated theory. That is to say,  
given \eqn{eq:ama}-(\ref{eq:amd}) it follows that
\be
\beta_{m^2} = \frac{1}{2}m_0^2\mu\frac{d}{d\mu}\left(\mu\frac{d}{d\mu}\gamma\right)
\label{eq:betam}
\ee
where in \eqn{eq:betam}, $\beta_{m^2}$ does not include $D$-tadpole
contributions (that is, at one loop it is given by \eqn{eq:bsoftc});  
the renormalisation of these is dealt with separately 
by $\beta_{\xi}$, as described in the last section. 

Note the  arbitrary  parameter $\kappa$ in \eqn{eq:amd}; its presence
means that  we can, in the \mssm, follow the usual procedure whereby 
the Higgs $B$-parameter is determined (along with the $\mu$-term) by 
the electroweak minimisation. How {\it natural\/} is this procedure  is
an obvious question, to which we will return later.

The approach to the \amsb\ tachyonic slepton problem 
that we will follow is based on the fact 
that \RG\ invariance is preserved if we replace 
$(m^2)^i{}_j$ in Eq.~(\ref{eq:amc}) by 
\be 
(\mbar^2)^i{}_j =  \frac{1}{2}m_0^2\mu\frac{d}{d\mu}\gamma^i{}_j
+k^{\prime}(\Ycal^{\prime})^i{}_j,
\label{eq:AE}
\ee
where $k^{\prime}$ is a constant and $\Ycal^{\prime}$ is a matrix 
satisfying 
\be
(\Ycal^{\prime})^i{}_l Y^{ljk} +  (\Ycal^{\prime})^j{}_l Y^{ilk} +
(\Ycal^{\prime})^k{}_l Y^{ijl} = 0
\label{eq:ypinv}
\ee
and
\be
\Tr \left[\Ycal^{\prime}C(R_{\alpha})\right] = 0,
\label{eq:ypanom}
\ee
in other words $\Ycal^{\prime}$ is a hypercharge
matrix corresponding
to a $U_1$ symmetry (which we shall denote $U_1^{\prime}$)
with no mixed anomalies 
with the \sm{} gauge group. This $U_1^{\prime}$ may in general be 
gauged, or a global symmetry. 

The \mssm{} (including right-handed neutrinos) admits 
two independent generation-blind anomaly-free $U_1$ symmetries. 
The possible charge assignments are shown in Table~1. 
\begin{table}
\begin{center}
\begin{tabular}{|c c c c c c|} \hline
$Q$ & $u^c$ & $d^c $ 
& $H_1$ & $H_2$ & $\nu^c$ \\ \hline
& & & & & \\ %\hline
$-\frac{1}{3}L$ & $-e-\frac{2}{3}L$  & $e+\frac{4}{3}L$ 
& $-e-L$ & $e+L$ & $-2L-e$ \\ %\hline
& & & & & \\ \hline
\end{tabular}
\caption{\label{anomfree}Anomaly free $U_1$ charges for arbitrary 
lepton doublet and singlet charges $L$ and $e$ respectively. 
$U_1^{SM}$ corresponds to $L=-1/2$ and $e=1$. }
\end{center}
\end{table}

Of course the $k^{\prime}\Ycal^{\prime}$ 
term in \eqn{eq:AE}{} corresponds in form to  a FI $D$-term; 
we shall assume that in fact the associated  $U_1^{\prime}$ 
gauge symmetry is broken at high energy and 
that the above contributions to the scalar masses are the only 
relic of this breaking that survive in the 
low energy effective field theory. 
That this is a perfectly natural scenario 
was demonstrated in \reference{amsbone}.

Now let us consider a possible FI term $\xi D$ associated with the \sm\
(or \mssm)  $U_1$, $U_1^{SM}$. Here $\xi$ is an independent parameter
respecting  all the symmetries of the \mssm; in the vast majority of
analyses  using, for example, \cmssm{} boundary conditions at gauge
unification, it  is assumed to be zero there. (For an exception, in
which $\xi$ is treated as  an extra independent parameter at low energy,
see  \reference{de Gouvea:1998yp}).    Working in the $D$-eliminated
formalism, the effect of  radiative generation of an FI term as we run
down to low scales is then  automatically taken care of by the term
added in \eqn{eq:delim} (and  corresponding terms at higher loops). If,
on the other hand we work  with the $D$-uneliminated formalism then
obviously if we assume  $\xi$ is zero at gauge unification then it is
calculable at low energies  using $\beta_{\xi}$ from
Eqs.~(\ref{eq:bxione},\ref{eq:bxitwo}). 
The resulting additional contributions to the masses 
from \eqn{eq:dpot}\ will of course lead to precisely the same results 
for the masses as obtained directly from the running of the masses 
using the $D$-eliminated formalism.

How large the radiatively generated $\xi$ is  depends
on the boundary conditions we assume for the scalar masses at  gauge
unification. Let us consider first the standard
\cmssm\ (or \msugra) picture. In that case it is clear that with 
the assumption of a common scalar mass at gauge unification, 
$\beta_{\xi}^{(1)}$ vanishes there because $U_1^{SM}$ is 
free of gravitational anomalies:
\be
\Tr[\Ycal]= 0. \label{gravnat}
\ee
Moreover, and
less obviously,   $\beta_{\xi}^{(1)}$ is in fact \RG\ invariant; 
that is, 
using \eqn{eq:bsoftc} in 
\eqn{eq:xizero}\ we find that 
\be
\Tr\left[\Ycal \beta_{m^2}^{(1)}\right] = 0
\label{eq:xizero}
\ee
where we denote the \sm{} hypercharge by $\Ycal$. 
This follows because $\Ycal$ naturally satisfies 
\eqn{eq:ypinv}, (with $\Ycal^{\prime}$ replaced by $\Ycal$):
\be
\Ycal^i{}_l Y^{ljk} +  \Ycal^j{}_l Y^{ilk} +
\Ycal^k{}_l Y^{ijl} = 0
\label{eq:yinv}\ee
(similarly for $h^{ijk}$) and anomaly cancellation, 
\be
\Tr \left[\Ycal C(R_{\alpha})\right] = 0.
\label{eq:yanom}
\ee

So for \cmssm\ boundary conditions, or indeed {\it any\/} boundary
conditions  such that $\Tr\left[\Ycal m^2\right]=0$ at gauge
unification, then, in the one-loop  approximation, $\xi$ is zero at low
energy if it is zero at gauge unification. (If we go beyond one loop
then a non-zero but quite small $\xi$ will be generated.)

We turn now to the \amsb{} scenario. Substituting 
\eqn{eq:AE}\  in \eqn{eq:bxione}  and \eqn{eq:bxitwo} 
we find that  up to two loops we can write 
\be
\lf\bxhat=g_1|m_0|^2\left(\mu\frac{d}{d\mu}
\Tr[\Ycal (\gamma-\gamma^2 )]
+2k^{\prime}\Tr[\Ycal\Ycal^{\prime}(1-2\gamma)]\right),
\ee
and since
gauge invariance and anomaly cancellation combined with
Eqs.~(\ref{beone:2}) and (\ref{betwo:2}) yield\cite{xione}
\be
\Tr[\Ycal \gamma^{(1)}] = \Tr[\Ycal(\gamma^{(2)}-(\gamma^{(1)})^2 )]=0,
\ee
this reduces to
\be
\lf\bxhat=
2k^{\prime}g_1|m_0|^2\Tr[\Ycal\Ycal^{\prime}(1-2\gamma)].
\label{eq:bxhat}
\ee
Thus in the {\it absence\/} of the $\Ycal^{\prime}$ term (i.e. with the 
unmodified mass solution of Eq.~(\ref{eq:amc})) an appreciable 
$U_1^{\sm}$ FI term will not be generated by the running, and 
Eq.~(\ref{eq:amc})  will therefore be \RG{} invariant. 
This was the conclusion of \reference{xithree}. 

Using \eqn{eq:AE}\ however, we obtain \eqn{eq:bxhat}, which 
is non-vanishing even at leading order unless we choose 
the charges $\Ycal^{\prime}$ so that  
\be
\Tr[\Ycal\Ycal^{\prime}] = 0.
\label{eq:kinmix}
\ee
This was in fact the choice made in \reference{jja}, the motive  there
being to  suppress kinetic mixing between the $U_1^{\sm}$  and the
$U_1^{\prime}$  gauge bosons (in that paper we considered a
$U_1^{\prime}$ broken at rather lower  energies). With such a
$U_1^{\prime}$, the $\Ycal^{\prime}$ charges   
$L$ and $e$ satisfy
\be
3L + 7e = 0
\label{eq:kincon}
\ee
so they are opposite in sign. Consequently 
\eqn{eq:AE} alone would not suffice to escape the tachyonic
slepton problem (if it held at low energy).  In \reference{jja} it was
shown, however,  that replacing \eqn{eq:AE} by
\be 
(\mbar^2)^i{}_j =  \frac{1}{2}m_0^2\mu\frac{d}{d\mu}\gamma^i{}_j
+k(\Ycal^{\sm})^i{}_j + k^{\prime}(\Ycal^{\prime})^i{}_j,
\label{eq:AEn}
\ee
(with $\Ycal^{\prime}$ charges satisfying \eqn{eq:kincon}) 
could do so. Now since we have shown above that an 
effective $U_1^{\sm}$ FI-term 
is in any event generated by \RG{} running, it is 
not a priori obvious that 
having simply \eqn{eq:AE}\ {\it at gauge unification\/} 
even with a $U_1^{\prime}$ with opposite $L,e$ charges won't work; 
however we may expect that the $U_1^{\prime}$ 
choice of \reference{jja}\ clearly will not do, precisely because of 
\eqn{eq:kinmix}; the generated $\xi$ for $U_1^{\sm}$ 
will be too small. We shall see that this is indeed the case. 
One might hope that it would be possible to 
choose, for example, $U_1^{\prime} \equiv U_1^{B-L}$; 
we shall see, however, that, with \eqn{eq:AE}, 
although the region of $(e,L)$ 
parameter space corresponding to an acceptable \sic\ spectrum 
{\it does\/} indeed include the possibility of $L < 0$, it permits 
neither  \eqn{eq:kinmix}{} nor $L + e = 0$, 
which would have corresponded to $U_1^{B-L}$.  

Let us now follow \reference{amsbone}\ by considering a theory with 
FI-type contributions associated with $U_1^{\prime}$, and  compare the
consequences of imposing \eqn{eq:AE} (and vanishing  FI term for
$U_1^{\sm}$) at  (i) gauge unification  (ii) a common \susy\ scale,
$M_{SUSY}$.  It should be clear from the above discussion that using 
the same values of $(e,L)$ in the two cases will not give rise to the 
same spectrum, because imposing it at gauge unification  (say) will give
rise to a non-vanishing  $U_1^{\sm}$ FI term at $M_{SUSY}$, and 
corresponding contributions to the sparticle masses.

It is easy to see, however, that precisely the same spectrum 
consequent on a particular choice of $(e,L)$ at at $M_X$ can be 
obtained by by using a {\it different\/} $(e,L)$ pair at $M_{SUSY}$
(with in each case no $U_1^{\sm}$ FI term). 
This is simply because we can write 
\bea
\mbar_L^2 &=&  m_L^2 - \frak{1}{2}k + k^{\prime}L = 
 m_L^2 + k^{\prime\prime}L^{\prime\prime}\nn
\mbar_{e^c}^2 &=&   m_{e^c}^2 + k + k^{\prime}e = 
m_{e^c}^2 + k^{\prime\prime}e^{\prime\prime}\nn 
\mbar_{Q}^2 &=&   m_{Q}^2 + \frak{1}{6}k + k^{\prime}Q =
m_{Q}^2 + k^{\prime\prime}Q^{\prime\prime} \quad \hbox{etc.},
\eea
where $k^{\prime\prime}Q^{\prime\prime} = 
-k^{\prime\prime}\frak{1}{3}L^{\prime\prime}$, etc.

Thus we can absorb the $U_1^{\sm}$ FI term generated by the running 
into a redefinition of the charges $(e,L)$. 

Note that the above remarks strictly apply  only if we evaluate the
spectrum at a common mass scale, $M_{SUSY}$.  Since in
\reference{amsbone}\  we systematically evaluated each  sparticle pole
mass at a renormalisation scale  equal to the pole mass itself, small
discrepancies were introduced. From now on we will always 
calculate spectra by running down from $M_X$, inputting $(e,L)$ 
(and zero for the $U_1^{\sm}$ FI term) there.

\section{\label{sec:original}The MSSM and the sparticle spectrum}

The \mssm{} is defined by the superpotential: 
\be W =  H_2 Q Y_t t^c + H_1
Q Y_b b^c  +  H_1 L Y_{\tau} \tau^c + \mu H_1 H_2  
\ee with soft
breaking terms: 
\bea L_{\rm SOFT} & = & \sum_{\phi} m_{\phi}^2\phi^*\phi
+ \left[m_3^2 H_1 H_2 + \sum_{i=1}^3\half M_i\lambda_i\lambda_i  + {\rm
h.c. }\right]\nn 
& + & \left[H_2 Q h_t  t^c  + H_1 Q h_b b^c  + H_1 L
h_{\tau}\tau^c + {\rm h.c. }\right] 
\eea 
where in general $Y_{t,b,\tau}$
and $h_{t,b,\tau}$ are  $3\times 3$ matrices. We work throughout in the
approximation that the Yukawa  matrices are diagonal, and neglect the
Yukawa couplings of the  first two generations.

The anomalous dimensions of the  
Higgses and 3rd generation matter fields are given (at one loop) by  
\bea
\lf\gamma_{H_1} & = & 3\lambda_b^2+\lambda_{\tau}^2-\frak{3}{2}g_2^2
-\frak{3}{10}g_1^2
,\nn
\lf\gamma_{H_2} & = & 3\lambda_t^2-\frak32g_2^2-\frak{3}{10}g_1^2
,\nn
\lf\gamma_{L} & = & \lambda_{\tau}^2-\frak32g_2^2-\frak{3}{10}g_1^2
,\nn
\lf\gamma_{Q} & = & \lambda_b^2+\lambda_t^2-\frak83g_3^2-\frak32g_2^2
-\frak{1}{30}g_1^2,\nn
\lf\gamma_{t^c} & = & 2\lambda_t^2-\frak83g_3^2-\frak{8}{15}g_1^2,\nn
\lf\gamma_{b^c} & = & 2\lambda_b^2-\frak83g_3^2-\frak{2}{15}g_1^2,\nn
\lf\gamma_{\tau^c} & = & 2\lambda_{\tau}^2-\frak65g_1^2,
\eea
where $\lambda_{t,b,\tau}$ are the third generation Yukawa couplings. 
For the first two generations we use the same expressions 
but without the Yukawa contributions.  The two and three loop results 
for the anomalous dimensions and the gauge $\beta$-functions 
may be found in \reference{fjj}.

The soft scalar masses are given by 
\bea
\mbar^2_Q & = & m^2_Q -\frak{1}{3}Lk^{\prime},\quad
\mbar^2_{t^c} = m^2_{t^c} -(\frak{2}{3}L +e)k^{\prime},\nn
\mbar^2_{b^c} & = & m^2_{b^c} +(\frak{4}{3}L+e)k^{\prime},\quad
\mbar^2_L =  m^2_L +Lk^{\prime},\nn
\mbar^2_{\tau^c} & = & m^2_{\tau^c} +ek^{\prime},
\quad \mbar^2_{H_{1,2}}   =  m^2_{H_{1,2}} \mp (e+L)k^{\prime},  
\label{eq:smasses}
\eea
(with similar expressions for the first two generations) 
where $m^2_Q$ etc are the pure anomaly-mediation contributions, for 
example:
\be
m_Q^2=\frak{1}{2}m_0^2\mu\frac{d}{d\mu}\gamma_Q 
= \frak{1}{2}m_0^2 \beta_i \frac{\pa}{\pa\lambda_i} \gamma_Q 
\label{eq:squarks}
\ee
(here $\lambda_i$ includes all gauge and Yukawa couplings) and 
$k^{\prime}$ is the  effective FI parameter.

The 3rd generation $A$-parameters are given by 
\bea
A_t &=-m_0(\gamma_Q+\gamma_{t^c}+\gamma_{H_2}),\nn
A_b &=-m_0(\gamma_Q+\gamma_{b^c}+\gamma_{H_1}), \nn
A_{\tau} &=-m_0(\gamma_L+\gamma_{\tau^c}+\gamma_{H_1})
\label{eq:apams}
\eea
and we set the corresponding first and second generation 
quantities to zero. 
The gaugino masses are given by 
\be
M_{\alpha} = m_0 \left(\frac{\beta_{g_{\alpha}}}{g_{\alpha}}\right), 
\quad\hbox{for}\quad {\alpha}=1,2,3.
\label{eq:inos}
\ee
The manner in which the scale of the 
effective FI parameter contributions $k^{\prime}L$ etc. 
to the sparticle masses can naturally be 
of the same  order as the anomaly mediation contributions 
when a $U_1^{\prime}$ is broken at high energies is explained 
in \reference{amsbone}{} and \reference{amsbtwo}.

Clearly these FI contributions depend on two parameters, 
$Lk^{\prime}$ and $ek^{\prime}$. For notational simplicity we will 
set $k^{\prime} = 1(\TeV)^2$ from now on. 

We begin by choosing input values for $m_0$, $\tan\beta$, $L$, $e$ 
and $\hbox{sign}\mu$ at $M_X$  and then we calculate the appropriate
dimensionless coupling input values  at the scale $M_Z$ 
by an iterative procedure involving the sparticle spectrum, 
and the loop corrections to $\alpha_{1\cdots 3}$, 
$m_t$, $m_b$ and $m_{\tau}$, 
as described in Ref.~\cite{bpmz}. 
We define gauge unification by the meeting point of $\alpha_1$ and $\alpha_2$. 
For the top quark pole mass we use $m_t = 170.9\GeV$. 

We then determine a given sparticle  pole mass  by running the
dimensionless couplings up to a certain scale chosen  (by iteration) to
be equal to the pole mass itself,  and then using
Eqs.~(\ref{eq:squarks}), (\ref{eq:apams}), (\ref{eq:inos}) and 
including full one-loop corrections from \reference{bpmz}, 
and two-loop corrections to the top quark mass~\cite{Bednyakov:2002sf}.

As in Ref.~\cite{jjk}, we have compared the effect of using one, two
and three-loop anomalous dimensions and $\beta$-functions in the
calculations. Note that when doing the three-loop calculation, we  use
in \eqn{eq:squarks}, for example, the three loop approximation  for
both $\beta_i$ and $\gamma_Q$, thus including some higher order 
effects. 

\begin{figure}[h]
\epsfysize= 4in
\centerline{\epsfbox{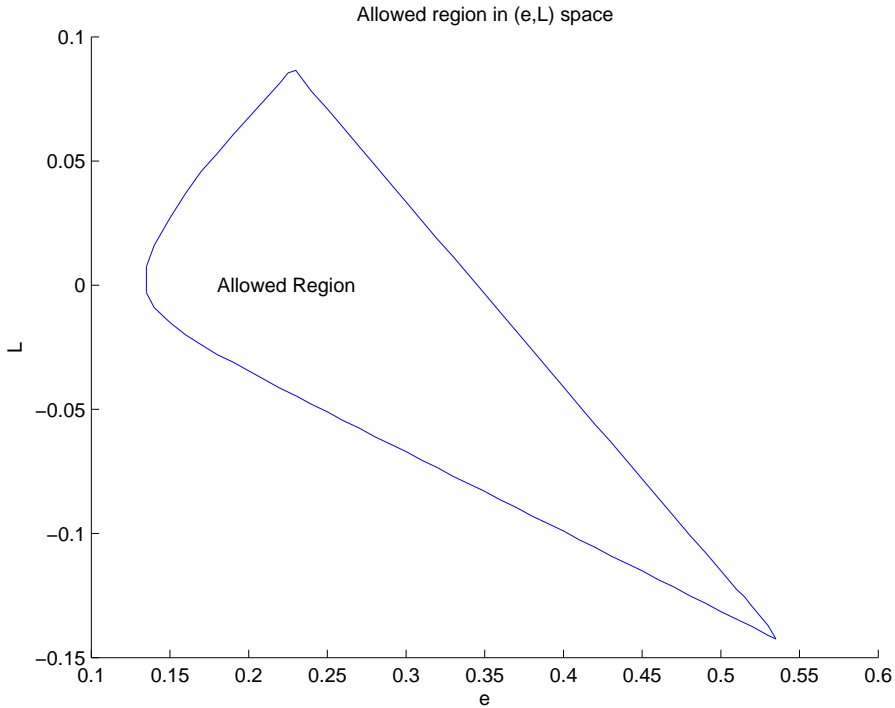}}
\in
\caption{\label{fig:Leregion}
The region of $(e,L)$ space corresponding to an acceptable 
electroweak vacuum, for $m_0 = 40\TeV$  
and $\tan\beta = 10$.}
\end{figure}
\medskip
\out

The allowed region in $(e,L)$ space for $\mu > 0$ and  $m_0 = 40\TeV$ 
corresponding to an acceptable vacuum 
is shown in Fig.~\ref{fig:Leregion}. To define the allowed region, 
we have imposed 
$m_{\tautil} > 82\GeV$, $m_{\nutil_{\tau}} > 49\GeV$ and $m_A > 90\GeV$. 
The region is to a good approximation triangular, with one 
side of the triangle corresponding to $m_A$ becoming too light 
(and quickly imaginary just beyond the boundary, with breakdown of 
the electroweak vacuum) and the other two sides to one of the sleptons 
(usually a stau) becoming too
light. 

Note that as we remarked earlier, the allowed region includes 
parts with $L < 0$. To understand this, consider, for example, 
the point $(e,L) = (0.35,-0.05)$. For this point we find that
\be
\Tr [\Ycal\Ycal^{\prime}] = 4.6.
\ee   
This is positive so from \eqn{eq:bxhat}{} we see that $\beta_{\xi}$ 
for the $U_1^{SM}$ FI term is positive at $M_X$. Since 
we are running {\it down\/} from $M_X$ it follows that 
a {\it negative\/} $\xi^{SM}$ is generated, and 
hence a {\it positive\/} contribution to $\mbar_L^2$, since 
the $U_1^{SM}$ charge of the lepton doublet is negative. Evidently 
the same reasoning means that we cannot have $e < 0$ at $M_X$, 
as we indeed see to be the case. 

Although $L < 0$ is allowed, it is easily seen that  we cannot, as we
mentioned earlier,  have either $L + e = 0$ (corresponding to
$U_1^{B-L}$) or  $3L + 7e = 0$  (corresponding to $\Tr
[\Ycal\Ycal^{\prime}] = 0$).

\begin{table}
\begin{center}
\begin{tabular}{| c | c | c | c |} \hline
 {\rm mass (GeV)}  &  1{\rm loop}   & 2{\rm loops} &  3{\rm loops}
 \\ \hline 
$ {\tilde g}$  &  925  &  900  & 897  \\ \hline
$ \ttil_1$  &  766 &  757 &  746 \\ \hline
$ \ttil_2$  &  502 &  500 &  487 \\ \hline
$ \util_L $  &  834 &  819 &  808 \\ \hline
$ \util_R $  &  774 &  766 &  753 \\ \hline
$ \btil_1$  &  724 & 712  &  702 \\ \hline   
$\btil_2$  & 956 & 946  & 936 \\ \hline
$ \dtil_L$  &  838  &  822 & 812\\ \hline
$\dtil_R$  &  965  & 955 & 946 \\ \hline
$\tautil_1$  & 267& 266&  266\\ \hline
$\tautil_2$  & 212 & 199 & 199\\ \hline
$ \etil_L$  & 262 & 261 & 262 \\ \hline
$\etil_R$  &  225 & 212 & 212\\ \hline
$\nutil_e $  & 250  &  249  & 249  \\ \hline
$\nutil_{\tau} $ &  248  & 247  & 247  \\ \hline
$\chi_1      $  &  106  & 131 & 131 \\ \hline
$\chi_2      $  & 354  &  362 &  362  \\ \hline
$\chi_3      $  &  569 &  593  &  585  \\ \hline
$\chi_4      $  &  580 &  604  & 596 \\ \hline
$\chi^{\pm}_1$  & 107 & 131  &  131  \\ \hline
$\chi^{\pm}_2$   & 577  &  601  &594 \\ \hline
$h      $  & 114 & 114   &  114  \\ \hline  
$H      $  & 333  & 373   &  361  \\ \hline
$A      $  & 333  & 373   & 361  \\ \hline
$H^{\pm}$ & 342 & 381  & 370 \\ \hline
$\chi^{\pm}_1-\chi_1$ (MeV)  &  {226}  & {235} & {237} \\ \hline
\end{tabular}
\caption{\label{table:spectrumA}
Mass spectrum for $m_t = 170.9\GeV$, $m_0 = 40\TeV$, $\tan\beta = 10$,
$L =  0$,  $e = 1/4$}
\end{center}
\end{table}

As an example of an acceptable spectrum, we give in Table~\ref{table:spectrumA} 
the results for $m_0 = 40\TeV, \tan\beta = 10, L = 0, e=1/4, 
\hbox{sign}\mu = +$
as derived using the one, two and three loop approximations 
for the anomalous dimensions and $\beta$-functions.  

This point in $(e,L)$ space is near the centre of  the allowed region
(see Fig.~\ref{fig:Leregion}).  As explained in the previous section,
the same spectrum would   be obtained to a  good approximation by
inputting parameters and  calculating pole masses at $M_Z$ with a
different pair of $(e,L)$  values, in this case $(e,L) \approx
(0.06,0.09)$. This point is near the centre of the allowed region in
Fig.~1 of \reference{amsbone}.   In Table~\ref{table:spectrumA},
however,  we give the masses with each calculated at a scale equal to
its pole mass. Therefore  as explained before, this means the whole
spectrum corresponds to choosing  the FI $U_1^{SM}$ term to be zero at
$M_Z$, but to a set of $(e,L)$ close to but each  differing  slightly
from $(0.06,0.09)$.

For the choice of parameters leading to Table~\ref{table:spectrumA}, we 
find that $\mu\sim 576\GeV$ and $B\sim (140\GeV)^2$, 
leading to $\kappa\sim 0.008$.  We see
that, aside from the little hierarchy problem associated  with the fact
that $\mu >> M_Z$, we have the problem of accounting for the  small
value of $\kappa$, and a degree of fine tuning between the two terms in 
\eqn{eq:amd}. As in \reference{amsbone}{} we find that to obtain 
a sufficiently high light CP-even Higgs mass, $m_h$ and an electroweak 
vacuum  we need to have 
$25 \gae  \tan\beta \gae 8$. 

For discussion of \amsb{} characteristic phenomenology 
the reader is referred to Refs.~\cite{lrrs}-\cite{amsbtwo}, and 
in particular \reference{ggw}.

\section{\label{GUT}$U_1$ and GUTs}

In the previous sections we have been assuming that our theory has 
gauge group $\Gcal_{SM} \otimes U_1^{\prime}$, broken to 
$\Gcal_{SM}$ at high energies. Let us now ask what modifications ensue 
if we ask for compatibility with a simple GUT embedding; for 
definiteness let us take $SU_5$, and imagine that our matter fields 
form a set of $n_f$ $(\fivebar + 10)$ multiplets as usual, and 
promote our Higgs multiplets to $n_h$ sets of $(5 + \fivebar)$. 
Then for compatibility with 
an $SU_5 \otimes U_1^{\prime}$ embedding we at once have the relations
\bea
Q &=& u^c = e\nn
d^c &=&  L   
\eea
and for $U_1^{\prime}$ invariance of the Yukawa terms
\bea
h_1 &=& -L -e\nn
h_2 &=&  -2e\nn
\nu^c &=&  2e - L.
\eea
Then the $SU_3^2\otimes U_1^{\prime}$, 
$SU_2^2\otimes U_1^{\prime}$ and $(U_1^{SM})^2\otimes U_1^{\prime}$ 
anomalies are all proportional to the quantity
\be
A_1 = (n_f - n_h)(L + 3e)
\ee
while the $(U_1^{\prime})^2\otimes U_1^{\sm}$
anomaly vanishes. The $(U_1^{\prime})^3$ anomaly is proportional 
to 
\be
A_3 = (L+3e)\left[5(n_f-n_h)( L^2 + 3 e^2)
-n_f(L + 3e)^2 \right]
\ee
while the $U_1^{\prime}-\hbox{gravitational}$ anomaly is proportional 
to
\be
A_G = (L+3e)(4n_f - 5 n_h).
\ee
Thus if $L + 3e = 0$ then the $\Gcal_{SM} \otimes U_1^{\prime}$ 
theory is anomaly-free for arbitrary $n_f, n_h$. This special case 
corresponds in fact to compatibility with 
the embedding $SO_{10}\supset SU_5\otimes U_1^{\prime}$ with 
each set of matter fields forming a 16 and each set of 
Higgs fields a 10 under 
$SO_{10}$. (Although $SO_{10}$ has complex representations 
they are all anomaly-free). Note the opposite sign charges for $L$ 
and $e$; we argued in Section~\ref{sec:amsbsol}{} 
that this does not preclude starting from $M_X$ with an FI term 
for such a $U_1^{\prime}$, but we shall see that the line $L+3e=0$ 
does not cross the allowed 
$(e,L)$ region for our class of models.
The other way to produce an anomaly-free theory is to first set 
$n_h = n_f$. Then $A_1=0$ while for $A_3$ and $A_G$ we have
\bea
A_3 &=& -n_f (L+3e)^3\nn
A_G &=& -n_f (L+3e)
\eea
so that we can obtain an anomaly-free theory by adding 
a further set of $n_f$  $\Gcal_{SM}$-singlet fields $N$, with 
charges $L + 3e$. The resulting charge assignments are shown in Table 3.
 
\begin{table}
\begin{center}
\begin{tabular}{|c c c c c c |} \hline
& & & & & \\ %\hline
$10$ & $\fivebar$ & $\nu^c $ 
& $H$ & $\overline{H}$ & N \\ 
 & & & & & \\ \hline
& & & & & \\ %\hline
$e$ & $L$  & $2e-L$ 
& $-2e$ & $-e-L$ & $L+3e$ \\ %\hline
& & & & & \\ \hline
\end{tabular}
\caption{\label{anomgutfree}Anomaly free $U_1$ symmetry for arbitrary 
lepton doublet and singlet charges}
\end{center}
\end{table}

This structure is compatible with $SU_5\otimes U_1^{\prime}$,  and can
be embedded in $E_6$,  when Table~\ref{anomgutfree}{} forms a $27$. 
(Recall that $E_6$ also has only anomaly-free representations).  If $L =
e$ we could have  $E_6 \supset SO_{10}\otimes U_1^{\prime}$,  
(with  Table~\ref{anomgutfree}{} forming a $16\oplus10\oplus1$ of $SO_{10}$),
or, as
explained above,   for $L = -3e$ we could have  $SO_{10}\supset
SU_5\otimes U_1^{\prime}$.  Another possibility is to have $L = 2e$, in
order that  $\nu^c$ have zero $U_1^{\prime}$ charge~\cite{king};
evidently this would have model-building  advantages if one  wants to
have a large mass for $\nu^c$ while  breaking $U_1^{\prime}$ at lower
energy.  Of course one sees easily that the cases $L=-3e$ and $L=2e$ are
equivalent  from a group theoretic point of view under the exchanges  $N
\leftrightarrow \nu^c$ and $\fivebar \leftrightarrow \Hbar$; obviously in the
latter case we could have an anomaly-free theory with $n_f$ sets of 
$(10,\Hbar,N)$ and $n_h$ sets of $(H,\fivebar)$.

Let us now suppose that, whatever the nature 
of the underlying theory, below gauge unification we have the
usual \mssm{} effective field theory, with three generations 
and a single pair of Higgs doublets (of course an explicit construction 
may lead to a more exotic low energy theory, but here we 
will confine ourselves to this possibility). We also assume FI contributions to 
the sparticle masses corresponding to our new $U_1^{\prime}$, thus 
instead of \eqn{eq:smasses}\ we have:
\bea
\mbar^2_Q & = & m^2_Q + ek^{\prime},\quad
\mbar^2_{t^c} = m^2_{t^c} +ek^{\prime},\quad
\mbar^2_{\tau^c} =  m^2_{\tau^c} +ek^{\prime},\nn
\mbar^2_{b^c} & = & m^2_{b^c} +Lk^{\prime},\quad
\mbar^2_L =   m^2_L +Lk^{\prime},\nn
\quad \mbar^2_{H_{1}}   & = &  m^2_{H_{1}} -(e+L)k^{\prime},\quad  
\quad \mbar^2_{H_{2}}   =   m^2_{H_{2}} -2ek^{\prime},  
\label{eq:smassesgut}
\eea
where $m^2_Q$ etc are again the pure anomaly-mediation contributions, 
and once again we set $k^{\prime}= 1$.

We can then compare the 
predicted sparticle spectrum with that obtained in the last 
section.  We may expect there to be differences, since 
evidently if we have both $(e,L) > 0$ it is now the case that 
both squarks and sleptons will have positive $(\hbox{mass})^2$ 
contributions. We calculate the spectrum as 
described in the previous section, running down from $M_X$; of course 
\RG{} invariance of the \amsb{} masses no longer holds because 
the effective field theory is no longer anomaly-free with respect to the 
$U_1^{\prime}$.

\begin{figure}[h]
\epsfysize= 4in
\centerline{\epsfbox{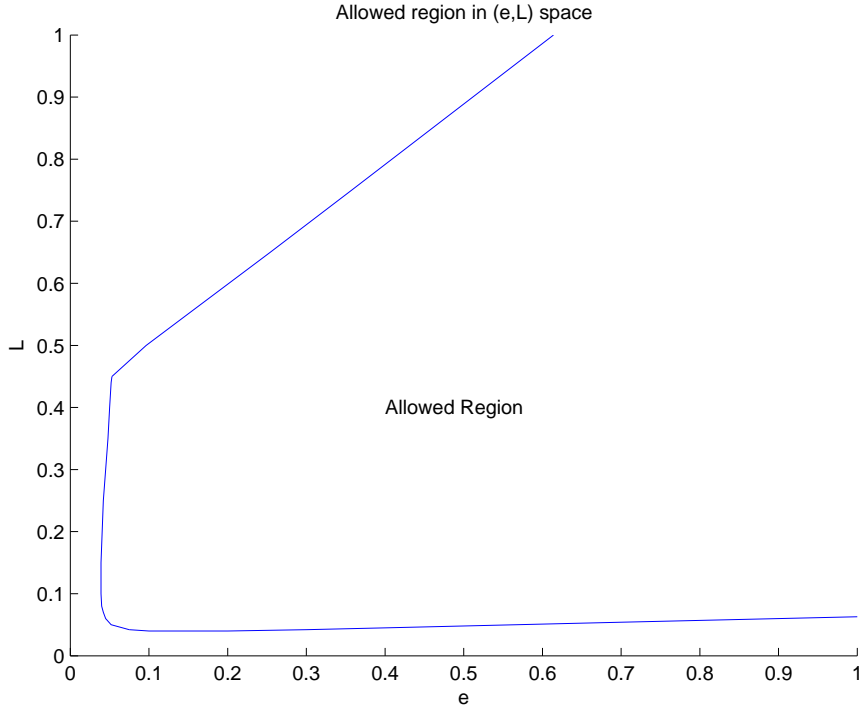}}
\in
\caption{\label{fig:gutfig}
The region of $(e,L)$ space corresponding to an acceptable 
electroweak vacuum, for $m_0 = 40\TeV$  
and $\tan\beta = 15$.}
\end{figure}
\medskip
\out

The allowed $(e,L)$ region with our new charge assignments is shown  in
\fig{fig:gutfig}. Comparing with \fig{fig:Leregion}, 
we see that the  most
dramatic difference is that increasing $(e,L)$ does 
not lead to loss of the electroweak vacuum 
as long as  $L \lae e + 0.4$. 
Of course increasing 
$(e,L)$ scales up the squark and slepton masses, 
$|m_{H_{1,2}^2}|$ and hence  the (Higgs)
$\mu$-parameter, thus increasing  the fine-tuning 
known as the little hierarchy problem. Other scenarios explored recently have  
also had this feature, for example split supersymmetry~\cite{split}, 
and the $G_2$ based model of \reference{gordy}. For a recent discussion 
of the little hierarchy problem see (for example) \reference{ibross}.

Another distinctive feature of the new charge assignment is that acceptable 
spectra are obtained with larger values of $\tan\beta$ than 
in section~\ref{sec:original}; here we find an upper limit of 
$\tan\beta = 43$. 

\begin{table}
\begin{center}
\begin{tabular}{| c | c | c | c |} \hline
 {\rm mass (GeV)}  &  1{\rm loop}   & 2{\rm loops} &  3{\rm loops}
 \\ \hline 
$ {\tilde g}$  &  966  &  940  & 938 \\ \hline
$ \ttil_1$  &  1063&  1047 &  1040 \\ \hline
$ \ttil_2$  &  936 &  923 & 917 \\ \hline
$ \util_L $  &  1103 & 1081 &  1073 \\ \hline
$ \util_R $  &  1102 &  1085 &  1077 \\ \hline
$ \btil_1$  &  1038& 974  &  1013\\ \hline   
$\btil_2$  & 993 & 1021  & 966 \\ \hline
$ \dtil_L$  &  1105  &  1084 & 1076\\ \hline
$\dtil_R$  &  1032 & 1014 & 1005 \\ \hline
$\tautil_1$  & 550   & 544  &  544  \\ \hline
$\tautil_2$  & 698   & 697  & 697  \\ \hline
$ \etil_L$  &  556  & 551 & 551 \\ \hline
$\etil_R$  &  698& 697  & 697 \\ \hline
$\nutil_e $  & 550  &  545  & 545  \\ \hline
$\nutil_{\tau} $  & 548  & 542  & 543  \\ \hline
$\chi_1      $  &  111  & 135 & 135  \\ \hline
$\chi_2      $  & 362  &  369 &  369  \\ \hline
$\chi_3      $  &  1204 &  1211  &  1207  \\ \hline
$\chi_4      $  & 1206 &  1213 & 1209  \\ \hline
$\chi^{\pm}_1$  & 111 & 135  &  136  \\ \hline
$\chi^{\pm}_2$   & 1207  &  1214  & 1210 \\ \hline
$h      $  & 115  & 115   &  115 \\ \hline  
$H      $  & 737  & 743   &  737  \\ \hline
$A      $  & 737  & 743   & 737  \\ \hline
$H^{\pm}$ & 742 & 748  & 742  \\ \hline
$\chi^{\pm}_1-\chi_1$ (MeV)  &  185  & 192 & 192 \\ \hline
\end{tabular}
\caption{\label{table:spectgut}
Mass spectrum for $m_t = 170.9\GeV$, $m_0 = 40\TeV$, $\tan\beta = 15$,
$L =  1/3$,  $e = 1/2$}
\end{center}
\end{table}

In Table~\ref{table:spectgut}\ we give results for the sparticle spectrum 
for a representative point in the allowed region. Of course $L=1/3$ and 
$e=1/2$ represent significant contributions to the squark squared masses, which 
are in any case already positive in \amsb, so it is not surprising that these 
masses are quite large for this point. Correspondingly the value of 
$\mu$ determined from electroweak minimisation is quite high  at around 
$1\TeV$.   

Both $L=e$ (corresponding to a potential $SO_{10}\otimes U_1^{\prime}$ 
embedding) and $L=2e$ (corresponding to zero $U_1^{\prime}$ charge for
$\nu^c$) are allowed; in the latter case we would need to have $e \lae
0.4$. In Table~\ref{table:spectgutb}\ 
we give results for the sparticle spectrum 
for $L = e = 0.1$, while in Table~\ref{table:spectgutc}\ 
we give results for the sparticle spectrum for $L = 2e = 0.1$.

\begin{table}
\begin{center}
\begin{tabular}{| c | c | c | c |} \hline
 {\rm mass (GeV)}  &  1{\rm loop}   & 2{\rm loops} &  3{\rm loops}
 \\ \hline 
$ {\tilde g}$  &  934  &  910  & 907  \\ \hline
$ \ttil_1$  &  858 &  847 &  838 \\ \hline
$ \ttil_2$  &  688 &  680 &  672 \\ \hline
$ \util_L $  &  908 &  891 &  881 \\ \hline
$ \util_R $  &  911 &  899 &  889 \\ \hline
$ \btil_1$  &  803 & 789  &  780 \\ \hline   
$\btil_2$  & 894 & 882 & 872  \\ \hline
$ \dtil_L$  &  911  &  894 & 885\\ \hline
$\dtil_R$  &  916 & 904 & 894  \\ \hline
$\tautil_1$  & 236   & 231  &  231  \\ \hline
$\tautil_2$  & 311   & 308  & 308   \\ \hline
$ \etil_L$  &  282  & 275 & 275 \\ \hline
$\etil_R$  &  282 & 281  & 281 \\ \hline
$\nutil_e $  & 270  &  263  & 263  \\ \hline
$\nutil_{\tau} $  & 266  & 259  & 259  \\ \hline
$\chi_1      $  &  109  & 133 & 134  \\ \hline
$\chi_2      $  & 358 &  365 &  365  \\ \hline
$\chi_3      $  &  820 &  833  &  828  \\ \hline
$\chi_4      $  &  826 &  839 & 834  \\ \hline
$\chi^{\pm}_1$  & 109& 134  &  134  \\ \hline
$\chi^{\pm}_2$   & 826  &  839  & 834 \\ \hline
$h      $  & 115  & 115   &  115 \\ \hline  
$H      $  & 623  & 635   &  629  \\ \hline
$A      $  & 624  & 636   & 629  \\ \hline
$H^{\pm}$ & 629 & 641  & 634  \\ \hline
$\chi^{\pm}_1-\chi_1$ (MeV)  &  192  & 199 & 200 \\ \hline
\end{tabular}
\caption{\label{table:spectgutb}
Mass spectrum for $m_t = 170.9\GeV$, $m_0 = 40\TeV$, $\tan\beta = 15$,
$L = e =0.1$}
\end{center}
\end{table}

\begin{table}
\begin{center}
\begin{tabular}{| c | c | c | c |} \hline
 {\rm mass (GeV)}  &  1{\rm loop}   & 2{\rm loops} &  3{\rm loops}
 \\ \hline 
$ {\tilde g}$  &  930 &  906  &  903 \\ \hline
$ \ttil_1$  &  828 &  818 &  809 \\ \hline
$ \ttil_2$  &  650 &  642 &  633 \\ \hline
$ \util_L $  &  880 &  864 &  854 \\ \hline
$ \util_R $  &  884 &  873 &  863 \\ \hline
$ \btil_1$  &  771 & 759  &  749 \\ \hline   
$\btil_2$  & 893 & 882 & 872 \\ \hline
$ \dtil_L$  &  883  &  868 & 857\\ \hline
$\dtil_R$  &  916 & 905 & 895 \\ \hline
$\tautil_1$  & 290   & 285 &  285  \\ \hline
$\tautil_2$  & 131   & 126  & 127  \\ \hline
$ \etil_L$  &  284  & 278 & 278 \\ \hline
$\etil_R$  &  165 & 162  & 162\\ \hline
$\nutil_e $  & 272  &  266  & 266  \\ \hline
$\nutil_{\tau} $  & 268  & 262  & 262  \\ \hline
$\chi_1      $  &  109  & 133 & 133\\ \hline
$\chi_2      $  & 358 &  365 &  365  \\ \hline
$\chi_3      $  &  759 &  774  &  768  \\ \hline
$\chi_4      $  &  766 &  780 & 775  \\ \hline
$\chi^{\pm}_1$  & 109& 133  &  134 \\ \hline
$\chi^{\pm}_2$   & 765  &  780  & 775 \\ \hline
$h      $  & 115  & 115   &  115  \\ \hline  
$H      $  & 585  & 599   &  591  \\ \hline
$A      $  & 585  & 599   & 592 \\ \hline
$H^{\pm}$ & 591 & 604  & 597  \\ \hline
$\chi^{\pm}_1-\chi_1$ (MeV)  &  195  & 203 & 203 \\ \hline
\end{tabular}
\caption{\label{table:spectgutc}
Mass spectrum for $m_t = 170.9\GeV$, $m_0 = 40\TeV$, $\tan\beta = 15$,
$L = 2e =0.1$}
\end{center}
\end{table}

\section{\label{sec:sumrules}Mass sum rules}

By taking appropriate linear combinations of squark and slepton
$(\hbox{masses})^2$  so that the $(e,L)$ contributions cancel  it is
straightforward  to derive a pair of interesting sum rules similar to
those we derived~\cite{jja,amsbone},  with the original charge
assignments of Section~\ref{sec:original}:
\bea
m_{\util_L}^2+m_{\dtil_L}^2-m_{\util_R}^2-m_{\etil_R}^2 & \approx &
0.8\left(m_{\gtil}\right)^2,\nn
m_A^2 +\sec 2\beta\left (m_{\etil_R}^2-m_{\etil_L}^2\right)
-2M_W^2 +\frak{5}{2}M_Z^2
& \approx &
0.5\left(m_{\gtil}\right)^2,\nn
m_{\btil_1}^2 + m_{\btil_2}^2 - m_{\tautil_1}^2 - m_{\tautil_2}^2
& \approx & 1.5\left(m_{\gtil}\right)^2,\nn
m_{\btil_1}^2 + m_{\btil_2}^2 - m_{e_L}^2 - m_{e_R}^2
& \approx & 1.5\left(m_{\gtil}\right)^2,\nn
m_{d_L}^2 + m_{d_R}^2 - m_{e_L}^2 - m_{e_R}^2
& \approx & 1.8\left(m_{\gtil}\right)^2.
\label{eq:sumc}
\eea

Although these sum rules are derived using the tree results for the
various masses they  hold reasonably well for the physical
masses.  The numerical coefficients on the RHS of \eqn{eq:sumc} are in
fact  slowly varying functions of $\tan\beta$; the above results
correspond to $\tan\beta = 15$.

\section{Conclusions}

The \amsb{}  scenario is an attractive alternative to (and 
distinguishable from) the \cmssm. With \amsb{} it is possible to  imagine a
theory where the only explicit scale in the effective  field theory is
the gravitino mass. An explicit  realisation of this idea was given in
\reference{amsbtwo}, where  the scale corresponding to the spontaneous
breaking of an additional  $U_1^{\prime}$ symmetry (needed to solve the 
tachyonic slepton problem) was generated by
dimensional transmutation. (This theory had the additional feature 
of a weakly coupled chiral matter multiplet whose fermionic component 
is a dark matter candidate). There is no obstacle in principle to
extending this idea  to a Grand Unified Theory, with the unification
scale similarly generated  by dimensional transmutation; this idea 
led us to consider the alternative charge assignments of 
section~\ref{GUT}. One possibility would be a variation of 
the inverted hierarchy 
model of Witten~\cite{Witten}, defined by the superpotential
\be
W=\lambda_1 \Tr (A^2 Y) + \lambda_2 X (\Tr A^2 - m^2)
\label{eq:witten}
\ee
where $A,Y$ are $SU_5$ adjoints and $X$ is a singlet. In its original
form,  \sy{} is broken spontaneously in the O'Raifertaigh manner; moreover 
$SU_5$ is broken to $SU_3\otimes SU_2\otimes U_1$, with the scale at which this occurs 
being unrelated to $m^2$, and  generated by dimensional
transmutation. Our variation would be to have $m^2 = 0$ in 
\eqn{eq:witten}, 
with the $SU_5$ breaking generated 
in similar fashion\footnote{A discussion of the $m^2 \to 0$ 
limit of Witten's model appears in \reference{Einhorn:1982pp}.}
but the \sy{} breaking provided instead  by
anomaly mediation. We will explore this model in more detail elsewhere.
 
We have shown that while a $U_1^{\prime}$ gauge symmetry  broken at high
energies can lead in a natural way to the  FI-solution to the \amsb{}
tachyonic slepton problem, care must be taken  with regard to the FI
term associated with $U_1^{SM}$. We have also shown how  an extension of
the minimal model permits a gauged $U_1^{\prime}$ compatible with  grand
unification, with, in this case, sparticle spectra characterised by both 
heavy squarks and heavy sleptons. 

\section*{\large Acknowledgements}

RH was supported by the STFC. DRTJ thanks Luminita Mihaila, 
Graham Ross and Stuart Raby for conversations. We also particularly thank 
Ting Wang for helpful correspondence.


\begin{thebibliography}{99}
% references
\bibitem{lrrs}
L.~Randall and R.~Sundrum,  {\it Nucl.\  Phys.}\  {\bf B 557} (1999) 79
\bibitem{glmr}G.F.~Giudice, M.A.~Luty, H.~Murayama and R.~Rattazzi,
{\it JHEP\/} {\bf 9812} (1998) 27
\bibitem{aprr}A.~Pomarol and  R.~Rattazzi, {\it JHEP \/} {\bf 9905} (1999) 013
\bibitem{ggw} 
T.~Gherghetta, G.F.~Giudice and  J.D.~Wells, {\it Nucl.\  Phys.}\  {\bf B 559} (1999) 27
\bibitem{mlrr}
M.A.~Luty and R.~Rattazzi, {\it JHEP \/} {\bf 9911} (1999) 001
\bibitem{clmp}
Z.~Chacko, M.A.~Luty, I.~Maksymyk and E.~Ponton, 
{\it JHEP \/} {\bf 0004} (2000) 001
\bibitem{kss}E.~Katz, Y.~Shadmi and Y.~Shirman, {\it JHEP \/} 9908 (1999) 015  
\bibitem{jjb}I.~Jack and D.R.T.~Jones, {\it Phys.\  Lett.}\  {\bf B 465} (1999) 148
\bibitem{jftm}J.L.~Feng and T.~Moroi,  {\it Phys.\ Rev.}\ {\bf D 61} (2000) 095004
\bibitem{gdk}G.D.~Kribs,  {\it Phys.\ Rev.}\ {\bf D 62} (2000) 015008
\bibitem{ssu}S.~Su, {\it Nucl.\  Phys.}\  {\bf B 573} (2000) 87
\bibitem{bmp} J.A.~Bagger, T.~Moroi and E.~Poppitz, {\it JHEP \/} 
0004 (2000) 009
\bibitem{rretal}R.~Rattazzi, A.~Strumia 
and J.~Wells, {\it Nucl.\  Phys.}\  {\bf B 576} (2000) 3
\bibitem{fpjw}F.E.~Paige and J.~Wells, hep-ph/0001249%
\bibitem{okada}N.~Okada,  {\it Phys.\ Rev.}\ {\bf D 65} (2002) 115009
\bibitem{Luty:2001zv}M.~Luty and R.~Sundrum,
 {\it Phys.\ Rev.}\ {\bf D 67} (2003) 045007
\bibitem{jja}I.~Jack and D.R.T.~Jones, {\it Phys.\  Lett.}\  {\bf B 482} (2000) 167
\bibitem{Arkani-Hamed:2000xj}
N.~Arkani-Hamed, D.E.~Kaplan, H.~Murayama and Y.~Nomura,
%``Viable ultraviolet-insensitive supersymmetry breaking,''
{\it JHEP \/} {\bf 0102} (2001) 041
\bibitem{Harnik:2002et}
R.~Harnik, H.~Murayama and A.~Pierce,
%``Purely four-dimensional viable anomaly mediation,''
{\it JHEP} {\bf 0208} (2002) 034
\bibitem{mwells}
B.~Murakami and J.D.~Wells,
%``Abelian D-terms and the superpartner spectrum of anomaly-mediated
%supersymmetry breaking,''
 {\it Phys.\ Rev.}\ {\bf D 68} (2003) 035006 
\bibitem{Kitano:2004zd}
R.~Kitano, G.~D.~Kribs and H.~Murayama,
%``Electroweak symmetry breaking via UV insensitive anomaly mediation,''
 {\it Phys.\ Rev.}\ {\bf D 70 } (2004) 035001
\bibitem{Ibe:2004gh}
M.~Ibe, R.~Kitano and H.~Murayama,
%``A viable supersymmetric model with UV insensitive anomaly mediation,''
 {\it Phys.\ Rev.}\ {\bf D 71} (2005) 075003
\bibitem{amsbone} R.~Hodgson,
I.~Jack, D.R.T.~Jones and G.G.~Ross, {\it Nucl.\ Phys.}\ {\bf B 728} (2005) 192
\bibitem{amsbtwo} D.R.T.~Jones and G.G.~Ross, 
{\it Phys.\ Lett.}\ {\bf B 642} (2006) 540
\bibitem{xione} I.~Jack and D.R.T.~Jones, 
{\it Phys.\ Lett.}\ {\bf B 473} (2000) 102
\bibitem{xitwo} I.~Jack, D.R.T.~Jones and S.~Parsons, 
{\it Phys.\ Rev.}\ {\bf D 62} (2000) 125022
\bibitem{xithree} I.~Jack and D.R.T.~Jones, 
{\it Phys.\ Rev.}\ {\bf D 63} (2001) 075010
\bibitem{de Gouvea:1998yp}
  A.~de Gouvea, A.~Friedland and H.~Murayama,
  %``Less minimal supersymmetric standard model,''
  {\it Phys.\ Rev.}\   {\bf D 59} (1999) 095008
 % [arXiv:hep-ph/9803481].
  %%CITATION = PHRVA,D59,095008;%%
\bibitem{fjj}P.M.~Ferreira, I.~Jack and D.R.T.~Jones, 
{\it Phys.\ Lett.}\ {\bf B 387} (1996) 80
%\bibitem{Erler:1999nx}
%J.~Erler and P.~Langacker, {\em Phys.~Rev, Lett.} { \bf 84} (2000) 212\semi
%%%CITATION = HEP-PH 9910315;%%
%D.A.~Demir, G.L.~Kane and T.T.~Wang,
%%``The minimal U(1)' extension of the MSSM,''
%hep-ph/0503290.
%%%CITATION = HEP-PH 0503290;%%
%\bibitem{mur}
%Y.~Kawamura, H.~Murayama and M.~Yamaguchi,
%\PRD 51 (1995) 1337\semi
%H.~Murayama, hep-ph/9503392
%\cite{de Gouvea:1998yp}
\bibitem{bpmz}D.M.~Pierce, J.A.~Bagger, K.T.~Matchev and R.J.~Zhang,
%``Precision corrections in the minimal supersymmetric standard model,''
{\it Nucl.\  Phys.}\  {\bf B 491} (1997) 3
\bibitem{Bednyakov:2002sf}
A.~Bednyakov et al, 
{\it Eur.\ Phys.\ J.\ } {\bf C 29} (2003) 87
%\bibitem{cdf}The CDF  Collaboration, the D0 Collaboration and the TEVEWWWG, 
%hep-ex/0507091
%\bibitem{hebb}T.~Hebbeker, {\it Phys.\  Lett.}\  {\bf B 470} (1999) 259
%\bibitem{hmrw} D.~Hooper, J.~March-Russell and S.M.~West,
%Phys.\  Lett.}\  {\bf B 605} (2005) 228
\bibitem{jjk} I.~Jack, D.R.T.~Jones, A.F.~Kord, 
{\it Ann.~Phys.\/} 316 (2005) 213
%\bibitem{jjyt} I.~Jack and D.R.T.~Jones {\it Nucl.\  Phys.}\  {\bf B 662} (2003) 63
%\bibitem{sven} S.~Heinemeyer, W.~Hollik and G.~Weiglein,
%                    {\em Eur.~Phys.~Jour.} { C 9} (1999) 343, 
%                    {\em Comp.~Phys.~Comm.} { 124} 2000 76\semi
%                    The code is accessible via
%                    {\tt www.feynhiggs.de} 
%\bibitem{Espinosa:2000df}
%J.~R.~Espinosa and R.~J.~Zhang,
%\NPB 586  (2000) 3;
%{\it JHEP} 0003 2000  026 
%\bibitem{Brignole:2002bz}
%A.~Brignole et al,
%\NPB 631 (2002) 195; {\it ibid\/} B643 (2002) 79
%\bibitem{martin}
%S.~P.~Martin, \PRD 67 (2003) 095012; {\it ibid\/} 67 (2003) 095012
\bibitem{king} S.F.~King, S.~Moretti and R.Nevzorov, 
{\it Phys.\  Lett.}\  {\bf B 650} (2007) 57
\bibitem{split}
  N.~Arkani-Hamed and S.~Dimopoulos,
  JHEP {\bf 0506} (2005) 073\semi
  %%CITATION = HEP-TH 0405159;%%
  G.F.~Giudice and A.~Romanino,
  {\it Nucl.\ Phys.}\ B {\bf 699} (2004) 65
  [Erratum-ibid.\ B {\bf 706} (2005) 65].
  %%CITATION = HEP-PH 0406088;%%  
\bibitem{gordy}
  B.S.~Acharya, K.~Bobkov, G.L.~Kane, P.~Kumar and J.~Shao,
  arXiv:hep-th/0701034.
  %%CITATION = HEP-TH/0701034;%%
%\cite{Ibanez:2007pf}
\bibitem{ibross}
  L.E.~Ibanez and G.G.~Ross,
  %``Supersymmetric Higgs and radiative electroweak breaking,''
  arXiv:hep-ph/0702046.
  %%CITATION = HEP-PH/0702046;%%
\bibitem{Witten}
  E.~Witten,
  %``Mass Hierarchies In Supersymmetric Theories,''
  {\it Phys.\ Lett.}\   {\bf B 105} (1981) 267 
  %%CITATION = PHLTA,B105,267;%%
\bibitem{Einhorn:1982pp}
  M.B.~Einhorn and D.R.T.~Jones,
  %``Scale Fixing By Dimensional Transmutation: Supersymmetric Unified Models
  %And The Renormalization Group,''
  {\it Nucl.\  Phys.}\  {\bf B211} (1983) 29.
  %%CITATION = NUPHA,B211,29;%%



%%%%%%%%%%%%%%%%%%%%%%%%%%%%%%%%%%%%%%%%%%%%%%%%%%%%%%%%%%%%%%%%%%%%%%%%%
\end{thebibliography}
\end{document}